\documentclass[%
 reprint,
superscriptaddress,
 amsmath,amssymb,
 aps,
]{revtex4-2}
\usepackage{url}
\usepackage{newtxtext, newtxmath}
\usepackage{textcase}
\usepackage{amsmath}
\usepackage{bm}
\usepackage{braket}
\usepackage{physics}
\usepackage{color}
\usepackage{graphicx}
\usepackage{hyperref}
\usepackage[version=3]{mhchem}
\usepackage{booktabs}
\hypersetup{
setpagesize=false,
bookmarksnumbered=true,
bookmarksopen=true,
colorlinks=true,
linkcolor=blue,
citecolor=blue,
urlcolor=blue
}

\begin{document}
\title{Collective mode excitations and simulated $L$-edge resonant-inelastic x-ray scattering spectra in\\
antiferromagnetic Ca$_2$RuO$_4$}

\author{Shunsuke Yamamoto}
\affiliation{
Department of Physics, Chiba University, Chiba 263-8522, Japan
}
\author{Yukinori Ohta}
\affiliation{
Department of Physics, Chiba University, Chiba 263-8522, Japan
}
\author{Koudai Sugimoto}
\affiliation{
Department of Physics, Keio University, Yokohama 223-8522, Japan
}

\begin{abstract}
Using the three-orbital Hubbard model, we investigate the low-energy excitation spectra in the antiferromagnetic phase of Ca$_2$RuO$_4$.
We calculate the dynamical susceptibilities in the low-energy region by the random phase approximation and find that the anisotropic dispersion of the transverse mode is in good agreement with the spectra recently reported by inelastic neutron scattering experiments.
By the fast-collision approximation, we simulate the resonant inelastic x-ray scattering (RIXS) spectra of the Ru $L_3$ edge from the dynamical susceptibilities.
We show that the dispersion of the transverse mode is clearly observed in the calculated RIXS spectra and that the polarization dependence of the incident x rays enables one to distinguish between the excitations of the in-plane transverse mode and out-of-plane transverse mode.
\end{abstract}

\maketitle
\section{Introduction}
In several $4d$ or $5d$ electron transition-metal compounds, the spin-orbit coupling (SOC) and electron correlation produce unique quantum states \cite{Cao2018,Takayama2021}.
In materials with $t_{2g}$ orbitals formed by a cubic crystal field, one electron has an effective orbital angular momentum of $\ell=1$, leading to various magnetic properties.
In $t_{2g}^5$ electron systems, for example, a layered perovskite \ce{Sr2IrO4} is a weak Mott insulator with a half-filled narrow isospin $j=1/2$ based band in the square lattice \cite{Kim2008, Arita2012PRL}.

The influence of SOC on $t_{2g}^4$ electron configuration systems has recently attracted much attention.
In SOC-dominated materials, the local spin $S=1$ and effective orbital angular momentum $L=1$ align antiparallel, resulting in a nonmagnetic state with total angular momentum $J=0$.
In the region where the crystal field due to distortion of octahedra and/or superexchange are sufficiently strong, the system may become magnetic.
The layered perovskite \ce{Ca2RuO4} that we focus on is in such a region, where a metal-insulator transition with a shortened $c$-axis occurs at $T_\mathrm{MI}\simeq 360$~K and an antiferromagnetic (AFM) transition occurs at $T_\mathrm{N}\simeq 110$~K \cite{Nakatsuji1997, Braden1998PRB,Alexander1999}.
The low-temperature phase of \ce{Ca2RuO4} is considered to be either an $S=1$ Heisenberg antiferromagnet, in which the electronic configuration is $d_{yz}^1d_{xz}^1d_{xy}^2$ \cite{Kunkemoller2015, Kunkemoller2017, Zhang2017, Zhang2020}, or an excitonic magnet, in which the triplons condense between $J=0$ and $J=1$ \cite{Khaliullin2013PRL,Akbari2014, Svoboda2017, Sato2019, Kaushal2017PRB, Kaushal2020PRB, Feldmaier2020}.

Study of collective mode excitations is essencial to understand properties of materials with long-range order.
In \ce{Sr2IrO4}, for example, $L$-edge resonant inelastic x-ray scattering (RIXS) spectra show the dispersion of magnons with energy transfer $\omega=0$ at momentum transfer $\bm q=(0,0)$ \cite{Kim2012, Bertinshaw2020PRB}, which is similar to that of well-known Heisenberg antiferromagnets such as \ce{La2CuO4} \cite{Braicovich2009}.
In the AFM phase of \ce{Ca2RuO4}, inelastic neutron scattering (INS) spectra show a gap in the spectrum at momentum transfer $\bm q = (\pi, \pi)$ and a dispersion maximum at $\bm q = (0, 0)$ \cite{Kunkemoller2015, Kunkemoller2017, Jain2017NP}, which is different from the spectrum expected from the magnon dispersion of a simple Heisenberg antiferromagnet.
Theoretical explanations made in previous studies include a Heisenberg-model description for $S = 1$, which incorporates single-ion anisotropy due to spin-orbit coupling \cite{Kunkemoller2015, Kunkemoller2017, Zhang2017, Zhang2020} and triplon condensation \cite{Jain2017NP, Khaliullin2013PRL, Akbari2014}.
These studies are based on the effective strong-coupling model.
However, more detailed models based on realistic electronic states of the system are needed to understand what kind of excitation structure is present in this material.
Recently, O $K$-edge \cite{Fatuzzo2015, Das2018} and Ru $L_3$-edge \cite{Gretarsson2019} RIXS has also been used to investigate the excitation spectra, which makes it possible to verify the collective excitation structure from various angles.

Using the three-orbital Hubbard model obtained by the band-structure calculation and applying mean-field approximation and random-phase approximation (RPA), we investigate the low-energy excitation spectra in the AFM phase of \ce{Ca2RuO4}.
We note that \ce{Ca_2RuO4} is known as a Mott insulator, in which we expect that the electronic correlation effect plays an important role.
However, while the approximations used in this study ignore a large part of the correlation effect,
we confirm that the calculated dispersion of the transverse mode is in good agreement with the spectra obtained by INS experiments.
Previous studies of \ce{Ca_2RuO4} within the mean-field approximation are also found in Refs.~\cite{Mohapatra2020JPCM, Mohapatra2021JPCM}.
By the fast-collision approximation, we simulate the RIXS spectra of the Ru $L_3$ edge from the dynamical susceptibility.
Indeed, preceding studies have shown that the RIXS spectra of low-energy excitations of AFM ground states in Mott insulators can be reproduced by the mean-field approximation plus RPA \cite{Igarashi2014PRB, Fidrysiak2020PRB, Fidrysiak2021PRB}.
We will show that in the calculated RIXS spectra, the dispersion of transverse mode is clearly observed.
We will also show that the polarization dependence of incident x rays enable one to distinguish between the excitations of the in-plane transverse mode and out-of-plane transverse mode.
We will also confirm that the high-energy RIXS intensity corresponding to excitations between the $d_{yz/xz}$ and $d_{xy}$ orbitals are significantly dependent on the angle of incidence, in agreement with previous studies \cite{Gretarsson2019}.

The rest of this paper is organized as follows.
In Sec.~\ref{sec:model_and_method}, the three-orbital Hubbard model with SOC is introduced as an phenomenological model for describing Ca$_2$RuO$_4$, together with the mean-field approximation.
We also introduce the RPA and analyze the dynamical magnetic susceptibility corresponding to INS spectra in this approximation.
In Sec.~\ref{sec:RIXS_spectrum}, RIXS spectra tuned for the Ru $L_3$ edge are calculated based on the fast-collision approximation.
The spectral characters in both low- and high-energy regions are discussed.
We summarize our results in Sec.~\ref{sec:conclusion}.

\section{Model and method}
\label{sec:model_and_method}

We introduce the three-orbital Hubbard model, including the SOC term as an effective model for Ca$_2$RuO$_4$.
By applying the mean-field approximation, we obtain the AFM ground state of the system.
The dynamical magnetic susceptibility is calculated in the RPA.

\subsection{Three-orbital model and mean-field approximation}

We consider the three-orbital Hubbard model on the square lattice with periodic boundary condition for modeling the $t_{2g}$ electrons of \ce{Ca_2RuO_4}.
We define that the $x$ and $y$ axes are parallel to the Ru bonds while the $z$ axis is perpendicular to the square lattice.
The lattice constant is set to be unity.
The Hamiltonian is written by $\mathcal{H} = \mathcal{H}_0 + \mathcal{H}_{\mathrm{SOC}} + \mathcal{H}_{\mathrm{int}}$, where $\mathcal{H}_0$ is the kinetic-energy term, $\mathcal{H}_{\mathrm{SOC}}$ is the SOC term, and $\mathcal{H}_{\mathrm{int}}$ is the interaction term.

The kinetic-energy term reads
\begin{align}
  \mathcal{H}_0
    = \sum_{\bm k,\sigma,\mu,\nu}\epsilon_{\mu,\nu}(\bm{k})
      c^\dag_{\bm k,\mu,\sigma} c_{\bm k,\nu,\sigma},
\end{align}
where $c^\dag_{\bm k,\mu,\sigma}$ is the creation operator of an electron with wave vector $\bm{k}$, orbital $\mu$ ($=yz, xz, xy$), and spin $\sigma$ ($=\uparrow,\downarrow$).
This term is estimated from the first-principles calculations.
First, we obtain the band structure, using the QuantumESPRESSO package \cite{Giannozzi2009JPCM, Giannozzi2017JPCM} with the revised Perdew-Burke-Ernzerhof generalized gradient approximation \cite{Perdew2008PRL} and the projector augmented-wave pseudopotential by Kresse and Joubert \cite{Joubert1999PRB, DalCorso2014CMS}.
The plane-wave cut-off energy is set to 60~Ry, and the $k$-point mesh on the $5\times 5 \times 2$ Monkhorst-Pack grid \cite{Monkhorst1976PRB} is used.
We use the crystal structure of \ce{Ca2RuO4} at 90~K~\cite{Porter2018}.
Then, we construct the maximally localized Wannier functions \cite{Nakamura2021CPC} for the energy window of $-1.9~\mathrm{eV} < E - E_F < 0.6~\mathrm{eV}$.

We consider only the nearest- and the next-nearest- neighbor hopping integrals.
The band structure is qualitatively the same as the original one.
The orbital-diagonal terms of $\epsilon_{\mu,\nu}(\bm{k})$ are given by
\begin{multline}
  \epsilon_{xy,xy}(\bm{k}) = -2t_{xy}^\mathrm{NN} \qty(\cos k_x + \cos k_y)\\
  - 4 t_{xy}^\mathrm{NNN} \cos k_x\cos k_y - \Delta,
\label{eq:e_xy-xy}
\end{multline}
\begin{align}
  \epsilon_{xz,xz}(\bm{k}) = -2t_{xz}^\mathrm{NN} \cos k_x,\quad
  \epsilon_{yz,yz}(\bm{k}) = -2t_{yz}^\mathrm{NN} \cos k_y,
\end{align}
whereas the orbital-off-diagonal terms are given by
\begin{align}
  \epsilon_{xz,xy}(\bm{k})
    = -2t' \cos k_x,\quad
  \epsilon_{yz,xy}(\bm{k})
    = -2t' \cos k_y.
\end{align}
$\Delta$ arises from the energy splitting due to compression of \ce{RuO_6} octahedra, and $t'$ arises from the rotation and tilt of the \ce{RuO6} octahedra.
The values of the parameters are $t_{xy}^\mathrm{NN} = 0.211$~eV, $t_{xz}^\mathrm{NN} = t_{yz}^\mathrm{NN} = 0.158$~eV, $t_{xy}^\mathrm{NNN} = 0.087$~eV, $t' = 0.052$~eV, and $\Delta = 0.240$~eV.

The effect of SOC and electron-electron interactions at each atomic site $i$ cannot be ignored in Ca$_2$RuO$_4$.
The SOC term is given by
\begin{align}
  \mathcal{H}_{\mathrm{SOC}}
    = \zeta \sum_{i} \bm{\ell}_{\mu,\nu} \vdot \bm{s}_{\sigma, \sigma'}
      c^\dag_{i, \mu, \sigma} c_{i, \nu, \sigma'},
\end{align}
where $\zeta$ is the strength of SOC, $\bm{s} = \bm{\sigma} / 2$ is the spin angular momentum with Pauli matrix $\bm{\sigma}$, and
\begin{align}
  \ell^x &=
  \begin{pmatrix}
    0 & 0 & 0 \\
    0 & 0 & i \\
    0 & -i & 0
  \end{pmatrix}
, \quad
  \ell^y =
  \begin{pmatrix}
    0 & 0 & -i \\
    0 & 0 & 0 \\
    i & 0 & 0
  \end{pmatrix}
  ,\notag\\ \quad
  \ell^z &=
  \begin{pmatrix}
    0 & i & 0 \\
    -i & 0 & 0 \\
    0 & 0 & 0
  \end{pmatrix}
\end{align}
is the orbital angular momentum for $t_{2g}$ electrons \cite{Sugano1970}.

The on-site interaction term is given by
\begin{align}
  \mathcal{H}_{\mathrm{int}}
    &= \frac{U}{2}\sum_{i,\mu,\sigma}
      c^\dag_{i\mu\sigma}c_{i\mu\sigma}c^\dag_{i\mu\bar{\sigma}}c_{i\mu\bar{\sigma}} \notag \\
    &+\frac{U'}{2}\sum_{i,\sigma,\sigma'}\sum_{\mu\ne\nu}
      c^\dag_{i\mu\sigma}c_{i\mu\sigma}c^\dag_{i\nu\sigma'}c_{i\nu\sigma'} \notag \\
    &-\frac{J}{2}\sum_{i,\sigma,\sigma'}\sum_{\mu\ne\nu}
      c^\dag_{i\mu\sigma}c_{i\mu\sigma'}c^\dag_{i\nu\sigma'}c_{i\nu\sigma} \notag \\
    &+\frac{J'}{2}\sum_{i,\sigma}\sum_{\mu\ne\nu}
      c^\dag_{i\mu\sigma}c^\dag_{i\mu\bar{\sigma}}c_{i\nu\bar{\sigma}}c_{i\nu\sigma},
\end{align}
where $U$, $U'$, $J$, and $J'$ are the intraorbital Coulomb interaction, interorbital one, Hund's rule coupling, and pair-hopping interaction, respectively.
We define $\bar{\uparrow} = \downarrow$ and $\bar{\downarrow} = \uparrow$.
We assume $J'=J$ and $U' = U-2J$, which are satisfied in an isolated ion \cite{Dagotto2001PR}.

We apply the mean-field approximation to the interaction terms to obtain the ground state.
We define the mean fields $\sum_{\bm{k}_0} \langle c_{\bm{k}_0, \mu, \sigma}^\dag c_{\bm{k}_0 + m\bm{Q}, \nu, \sigma'} \rangle$ with ordering vector $\bm{Q}$ for all combinations about orbital and spin, where $\bm{k}_0$ is the wave vector in the reduced Brillouin zone and $m$ is an integer.
In this paper, we assume a checkerboard-type order, i.e., $\bm{Q} = (\pi, \pi)$ and $m = 0, 1$.
The diagonalized mean-field Hamilitonian is written as
\begin{align}
  H^\mathrm{MF}
    = \sum_{\bm{k}_0, \epsilon}
      E_{\bm{k}_0, \epsilon}
      \gamma^\dag_{\bm{k}_0, \epsilon} \gamma_{\bm{k}_0, \epsilon},
\end{align}
where $\gamma_{\bm{k}_0,\epsilon}$ is the canonical transformation of annihilation operator satisfying $c_{\bm{k}_0+m\bm{Q},\mu,\sigma} = \sum_{\epsilon} \psi_{\mu,\sigma, m; \epsilon}(\bm{k}_0) \gamma_{\bm k_0,\epsilon}$, and $E_{\bm{k}_0, \epsilon}$ is the single-particle energy with band index $\epsilon$.
In this paper, we assume the absolute zero temperature and determine the Fermi energy $E_{\mathrm{F}}$ from the constraint that the number of particles per site equals to 4.
Using $100\times100$ meshes in the reduced Brillouin zone, we solved the mean-field equations self-consistently to calculate the order parameters, as discussed below.

\subsection{Magnetic moment}

The magnetic moment is obtained from the sum of the spin and orbital angular momentum, i.e.,
\begin{align}
  M^{\alpha} = 2 S^{\alpha} + L^\alpha,
\end{align}
where $L^\alpha = \ell^{\alpha} \otimes I_{2}$ and $S^{\alpha} = I_{3} \otimes s^{\alpha}$, and $I_{n}$ is the identity matrix of size $n$.
Using the creation and annihilation operators of electrons, the corresponding magnetic-moment operator with momentum $\bm{q}$ along $\alpha$ direction is expressed by
\begin{align}
  \hat{M}_{\bm{q}}^{\alpha}
    = 2 \hat{S}_{\bm{q}}^{\alpha} + \hat{L}_{\bm{q}}^\alpha
    = \sum_{\bm{k}}
      \bm{c}_{\bm{k}}^\dag M^{\alpha} \bm{c}_{\bm{k} + \bm{q}},
\label{eq:Mq}
\end{align}
where $\bm{c}_{\bm{k}} = (c_{\bm{k}, yz,\uparrow}, c_{\bm{k}, xz,\uparrow}, c_{\bm{k}, xy,\uparrow}, c_{\bm{k}, yz,\downarrow}, c_{\bm{k}, xz,\downarrow}, c_{\bm{k}, xy,\downarrow})^{t}$.

We calculate the average value of the magnetic moment by setting $\zeta = 0.15$~eV, $U = 2.0$~eV and $J=0.47$~eV, which are comparable to the ones used in previous theoretical studies \cite{Mizokawa2001, Gorelov2010PRL, Feldmaier2020}.
We find that, under the mean-field approximation, the AFM order with ordering vector $\bm{Q}$ is stable, and the spin moment and orbital moment are parallel.
The expectation value of each component is $\ev{\hat{S}^b_{\bm{Q}}} = 0.956$~$\mu_{\mathrm{B}}$, $\ev{\hat{S}^z_{\bm{Q}}} = 0.103$~$\mu_{\mathrm{B}}$, $\ev{\hat{L}^b_{\bm{Q}}} = 0.620$~$\mu_{\mathrm{B}}$, and $\ev{\hat{L}^z_{\bm{Q}}} = 0.072$~$\mu_{\mathrm{B}}$, and the components along $a$ axis are zero, where $a$ and $b$ axes are along $(1, -1)$ and $(1, 1)$ direcrions, respectively.
Thus, the magnetic moment lies in the plane nearly parallel to the $b$ axis.
The calculated magnetic moment 2.55~$\mu_{\mathrm{B}}$ is higher than the experimental value 1.3~$\mu_{\mathrm{B}}$ \cite{Braden1998PRB}.
This is due to the fact that angular momentum fluctuations are neglected in the mean-field approximation; a similar overestimation is also found in the previous studies using the mean-field approximation \cite{Mizokawa2001}.

 \begin{figure}
  \centering
  \includegraphics[width=1.0\columnwidth]{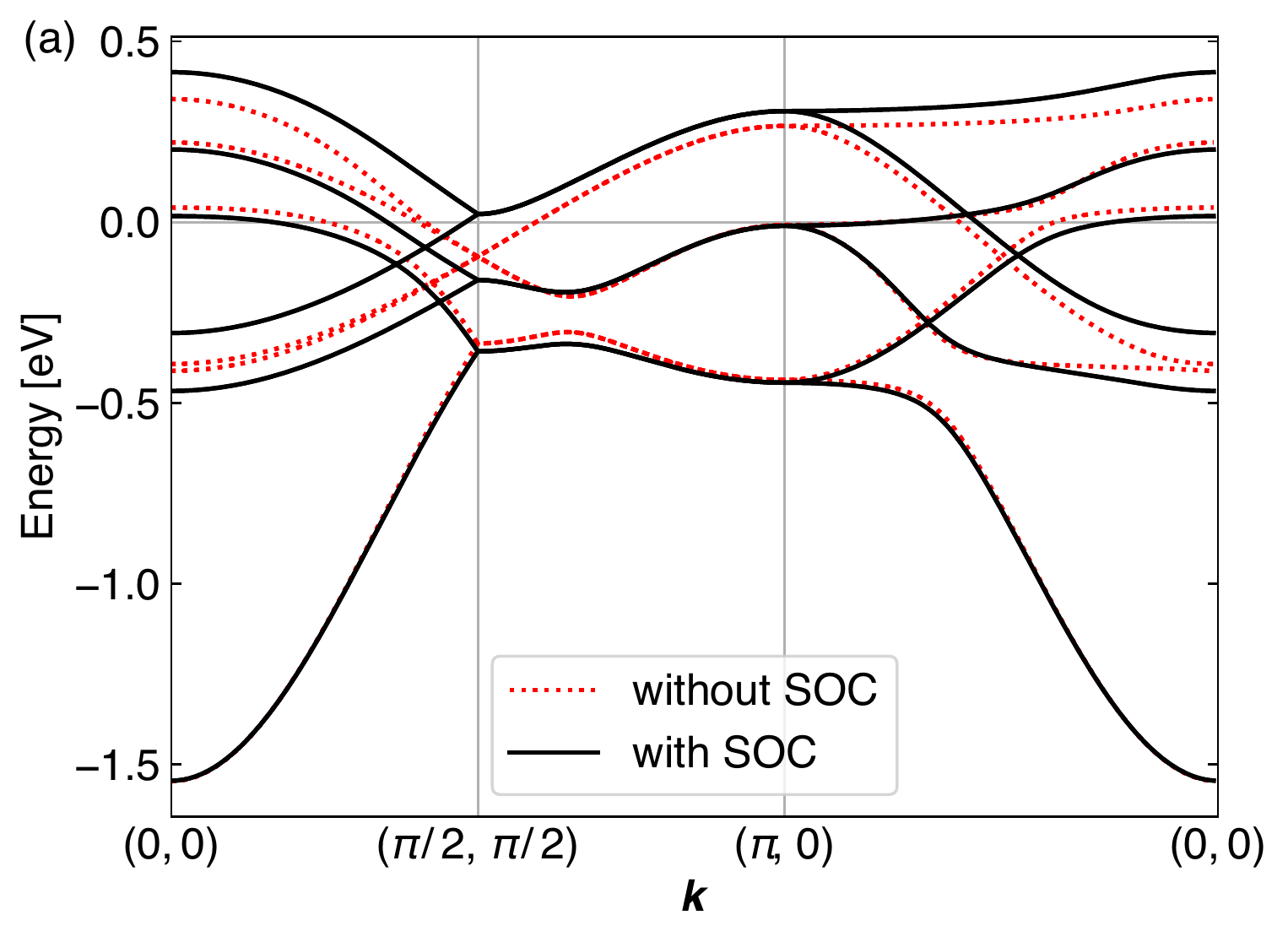}\\
  \includegraphics[width=1.0\columnwidth]{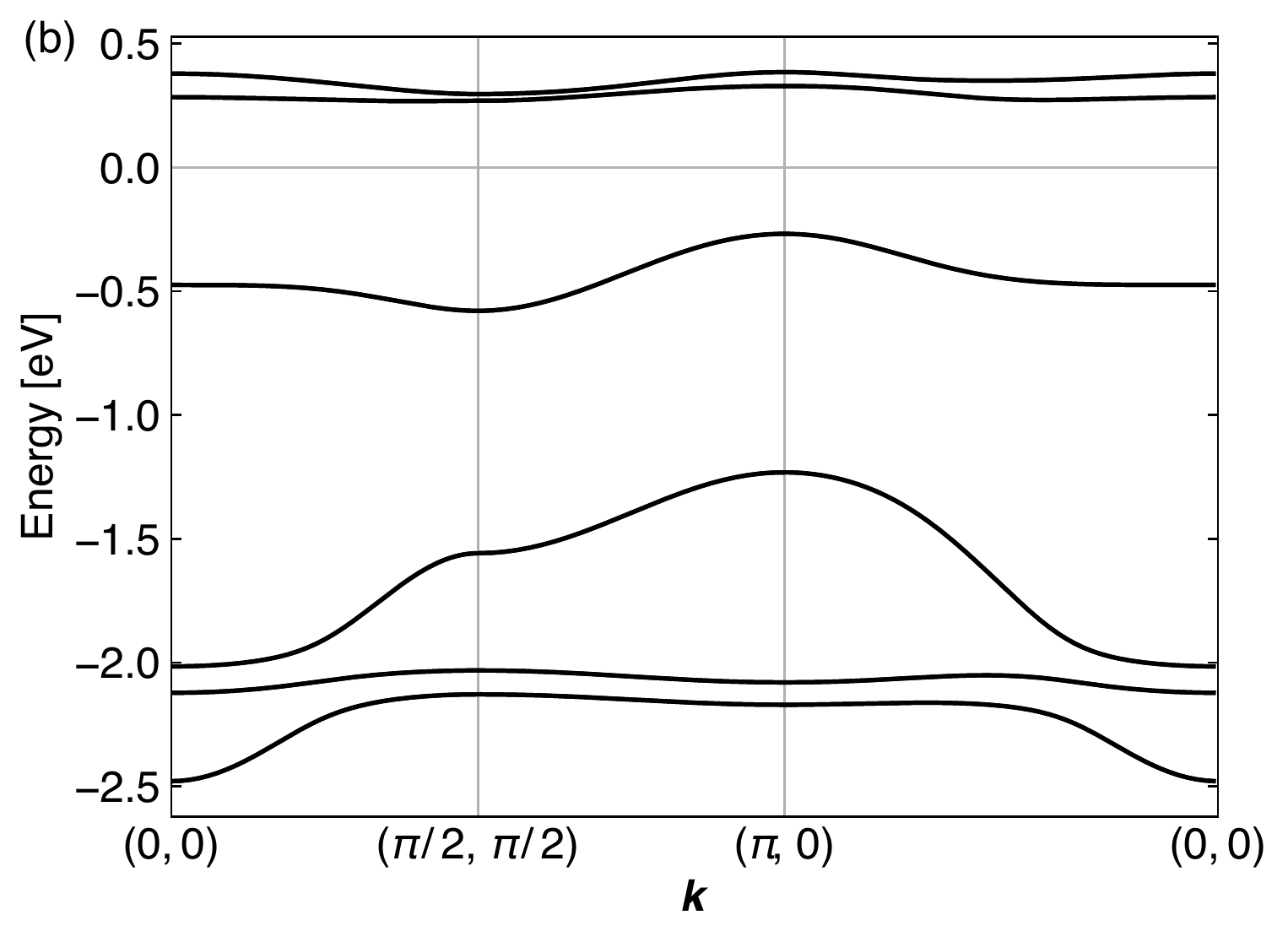}
  \caption{
    (a) Energy-band dispersions without the interaction term ($U=J=0$).
    The magnitude of the SOC is set to be $\zeta = 0.15$~eV.
    The dispersions without the SOC term ($\zeta = 0$) is also illustrated by the dotted red lines.
    (b) Energy-band dispersions obtained in the mean-field approximation with $U=2.0$~eV and $J = 0.47$~eV.
    The system is in the AFM state.
    In each figure, the energy bands are plotted in the reduced Brillouin zone.
  }\label{band}
\end{figure}

Figure \ref{band}(a) shows the energy dispersion without the interaction term ($U=J=0$).
Without the SOC ($\zeta=0$, dotted red lines), the upper four bands, which are degenerate at $(\pi/2,\pi/2)$, come from the $d_{yz/xz}$ orbitals.
These bands are lifted by introducing the SOC term ($\zeta = 0.15$~eV, solid black lines).
Introducing the interaction term ($U=2.0$~eV and $J=0.47$~eV) and assuming the AFM order in the mean-field approximation, the system becomes fully insulating, as shown in Fig.~\ref{band}(b).
There is one slow-dispersed band around $-0.5$~eV, two slow-dispersed bands around $-2.0$~eV, and one fast-dispersed band extended between $-2.5$~eV and $-1.2$~eV, which are consistent with the angle-resolved photoemission spectroscopy experiment \cite{Sutter2017}.

\subsection{Random-phase approximation}

The dynamical susceptibility of a multiorbital system is in general written as
\begin{multline}
  \chi_{u, v}
    \qty(\bm{q}_1,\bm{q}_2, \omega)
    = \frac{i}{N} \sum_{\bm{k}_1,\bm{k}_2} \int_0^\infty \dd{t} e^{i \omega t}\\
    \times \ev{\comm{c_{\bm{k}_1,\kappa, \sigma_1}^\dag(t)c_{\bm{k}_1 + \bm{q}_1,\lambda, \sigma_2}(t)}
    {c_{\bm{k}_2 + \bm{q}_2,\mu, \sigma_3}^\dag (0) c_{\bm{k}_2, \nu, \sigma_4} (0)}}\label{def_chi},
\end{multline}
where $N$ is the number of $\bm k$ points in the Brillouin zone, and $c_{\bm k,\mu,\sigma}(t)$ is the Heisenberg representation of $c_{\bm k,\mu,\sigma}$.
We define $u = (\kappa, \sigma_1; \lambda, \sigma_2)$ and $v = (\mu, \sigma_3; \nu, \sigma_4)$.
Hereafter, we consider the case $\bm{q}_1 = \bm{q} + l_1 \bm{Q}$ and $\bm{q}_2 = \bm{q} + l_2 \bm{Q}$.
The bare susceptibility is given by
\begin{align}
  &{\chi_{0}}_{u,v} (\bm{q} + l_1 \bm{Q}, \bm{q} + l_2 \bm{Q}, \omega)
  \notag\\
    &= \frac{1}{N}\sum_{\bm p_0,m,n,\epsilon,\epsilon'}\frac{f(E_{\bm p_0 + \bm q, \epsilon})-f(E_{\bm p_0, \epsilon'})}{E_{\bm p_0 + \bm q, \epsilon} - E_{\bm p_0 + \bm q, \epsilon'} - (\omega + i\eta)}
  \notag \\
  & \quad \times\psi_{\lambda, \sigma_2, m + l_1; \epsilon} (\bm{p}_0 + \bm{q})
  \psi^{*}_{\mu,\sigma_3, m + n + l_2;\epsilon}(\bm{p}_0 + \bm{q}) \notag \\
  & \quad \times\psi_{\kappa, \sigma_1, m;\epsilon'}^*(\bm{p}_0)
  \psi_{\nu, \sigma_4, m + n ;\epsilon'} (\bm{p}_0),
\end{align}
where $f(E)$ is the Fermi distribution function and the summation with respect to $\bm p_0$ runs over the reduced Brillouin zone.
\begin{table*}
  \centering
  \caption{Nonzero elements of $V_{v,u}$.}
  \label{Vele}
  \begin{tabular}{cccccc}
  \hline
  &$\sigma_1 = \sigma_2=\sigma_3=\sigma_4$
  &\quad&$\sigma_1=\sigma_2\ne\sigma_3=\sigma_4$
  &\quad&$\sigma_1=\sigma_4\ne\sigma_2=\sigma_3$\\
  \hline
  $\mu=\nu=\kappa=\lambda$&---& &$-U$& &$U$\\
  $\mu=\kappa\ne\nu=\lambda$&---& &$-J$& &$J$\\
  $\mu=\nu\ne\kappa=\lambda$&$-U + 3J$& &$-U + 2J$& &$J$\\
  $\mu=\lambda\ne\nu=\kappa$&$U - 3J$& &$-J$& &$U-2J$\\
  \hline
  \end{tabular}
\end{table*}
\begin{figure}
    \includegraphics[width=1.0\columnwidth]{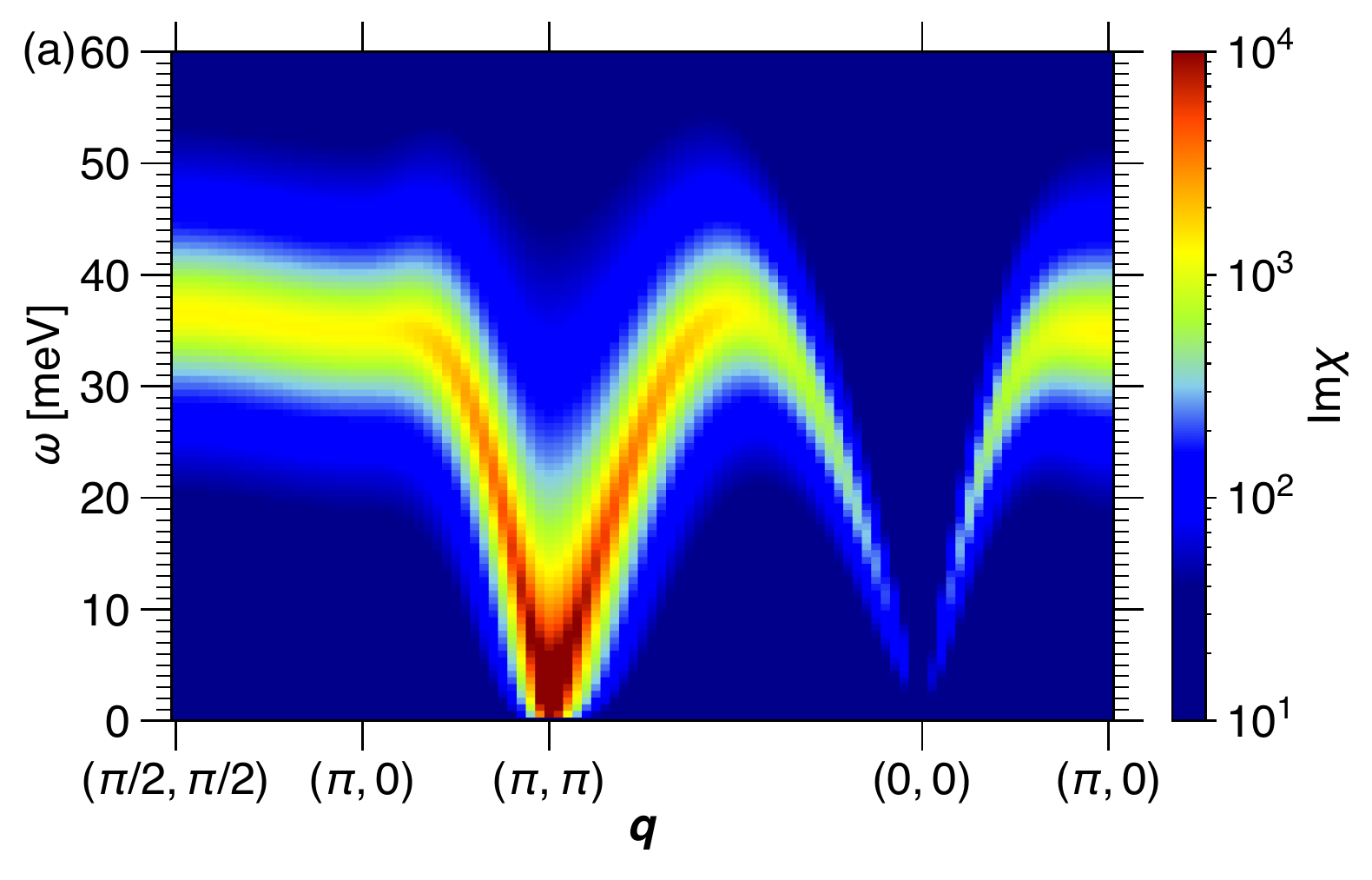}\\
    \includegraphics[width=1.0\columnwidth]{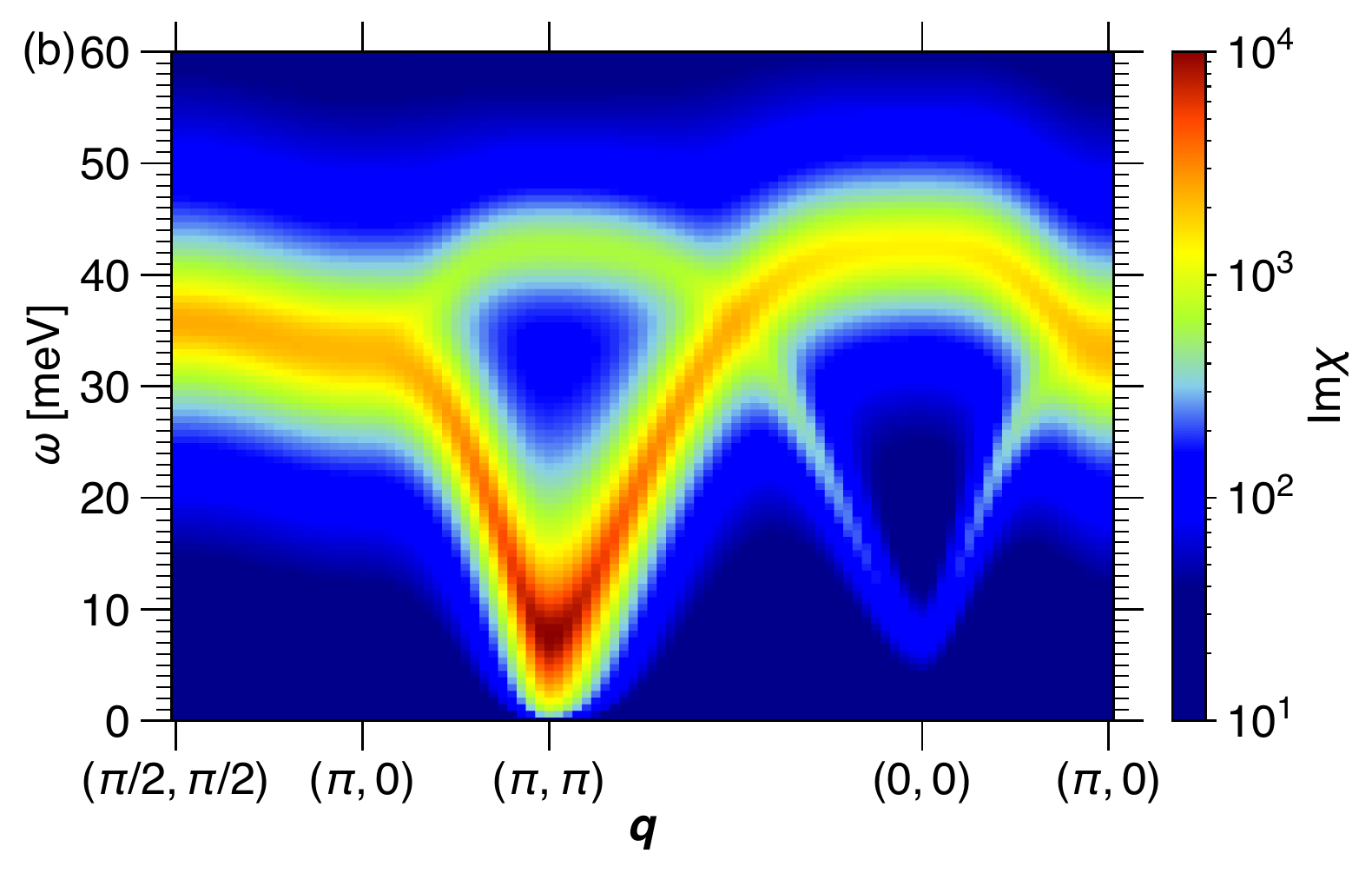}
  \caption{
   Calculated imaginary part of the dynamical magnetic susceptibility (a) without SOC ($\zeta = 0$) and (b) with SOC ($\zeta = 0.15$~eV).
We set the broadening parameter $\eta = 3$~meV.}
  \label{fig:mag_sus}
\end{figure}
We calculate the dynamical susceptibility in the multi-orbital RPA \cite{Sugimoto2013PRB}; i.e.,
\begin{multline}
  \chi_{u,v} \qty(\bm{q} + l_1 \bm{Q}, \bm{q} + l_2 \bm{Q}, \omega)
    = {\chi_{0}}_{u,v} \qty(\bm{q} + l_1 \bm{Q}, \bm{q} + l_2 \bm{Q}, \omega)
  \\
    + \sum_{u',v',l'}
    {\chi_{0}}_{u,v'} (\bm{q} + l_1 \bm{Q}, \bm{q} + l' \bm{Q}, \omega)
    V_{v', u'}
    \chi_{u', v} (\bm{q} + l' \bm{Q}, \bm{q} + l_2 \bm{Q}, \omega)
\end{multline}
with the interaction matrix $V$ listed in Table \ref{Vele}.
We abbreviate Eq. (\ref{def_chi}) as $\chi_{u,v} (\bm{q},\omega)$ when $\bm{q}_1 = \bm{q}_2 = \bm{q}$.

\subsection{Dynamical magnetic susceptibility}
\label{subsec:dynamical_magnetic_susceptibility}

Before calculating the RIXS spectra, we investigate the dynamical magnetic susceptibility, whose imaginary part directly corresponds to the excitation spectra observed in INS experiments.
The $\alpha$ component of the dynamical magnetic susceptibilities is given by
\begin{equation}
  \chi^{\alpha\alpha} \qty(\bm{q}, \omega)
    = \frac{i}{N} \int_0^\infty \dd{t}
      e^{i\omega t}\ev{\comm{M^{\alpha}_{\bm{q}} (t)}{M^{\alpha}_{-\bm{q}}(0)}},
\end{equation}
where the magnetic moment is defined in Eq. (\ref{eq:Mq}).

We carry out the calculation of the dynamical susceptibility with $N = 50 \times 50$ meshes for $k$-space integration.
Figure~\ref{fig:mag_sus} shows the imaginary part of the total dynamical magnetic susceptibility $\Im \chi \qty(\bm{q}, \omega) = \Im \sum_{\alpha} \chi^{\alpha \alpha} \qty(\bm{q}, \omega)$.
In Fig.~\ref{fig:mag_sus}(a), we consider the case where the SOC term in the Hamiltonian is neglected, i.e., $\zeta = 0$.
In this case, we observe the strong intensity at $\bm{q} = (\pi, \pi)$, which corresponds to the transverse-mode excitation.
Since the system is free from the SOC, the quantization axis of the antiferromagnetically ordered spins can be chosen in an arbitrary direction, and therefore the excitation gap of the spin-transverse mode closes at $\bm{q} = (\pi, \pi)$.
Also, the peak position of the excitation goes to $\omega = 0$ at $\bm{q} = (0, 0)$.
This behavior is the same as the spin-wave dispersion of the AF Heisenberg model \cite{Jain2017NP} or the single-band Hubbard model at half-filling in the AF state \cite{Ichioka2001JPSJ}.
Thus, the excitation spectra can be interpreted to be the usual spin-wave dispersion of the AF state when the SOC is absent.

Now, we turn on the SOC term.
The magnitude of the SOC is set to be $\zeta = 0.15$~eV.
The imaginary part of the dynamical susceptibility, in this case, is shown in Fig.~\ref{fig:mag_sus}(b).
We find that the strong intensity appears at $\bm{q} = (\pi, \pi)$, as in the case of $\zeta = 0$, but the peak position locates at finite frequency; i.e., the collective excitation is gapped at this point.
This is because the finite $\zeta$ and $t'$ in the Hamiltonian cooperatively introduce the magnetic anisotropy to the system.
Apart from the case without the SOC, the dispersion of the collective excitation reaches a maximum at $\bm{q} = (0,0)$.
This behavior resembles the spin-wave dispersion in a typical XY model \cite{Jain2017NP}.
By decomposing the dynamical magnetic susceptibility according to the direction of magnetization, we find that this excitation dispersion originates from the in-plane transverse component.
The out-of-plane transverse component shows a maximum at $\bm{q} = (\pi, \pi)$ and a minimum at $\bm{q} = (0, 0)$, which can be seen as a comparably weak intensity in Fig.~\ref{fig:mag_sus}(b).
The in-plane and out-of-plane transverse modes are degenerate along $\bm q = (\pi/2,\pi/2)$ to $(\pi,0)$.

\begin{figure}
  \begin{center}
    \centering
    \includegraphics[width=1.0\columnwidth]{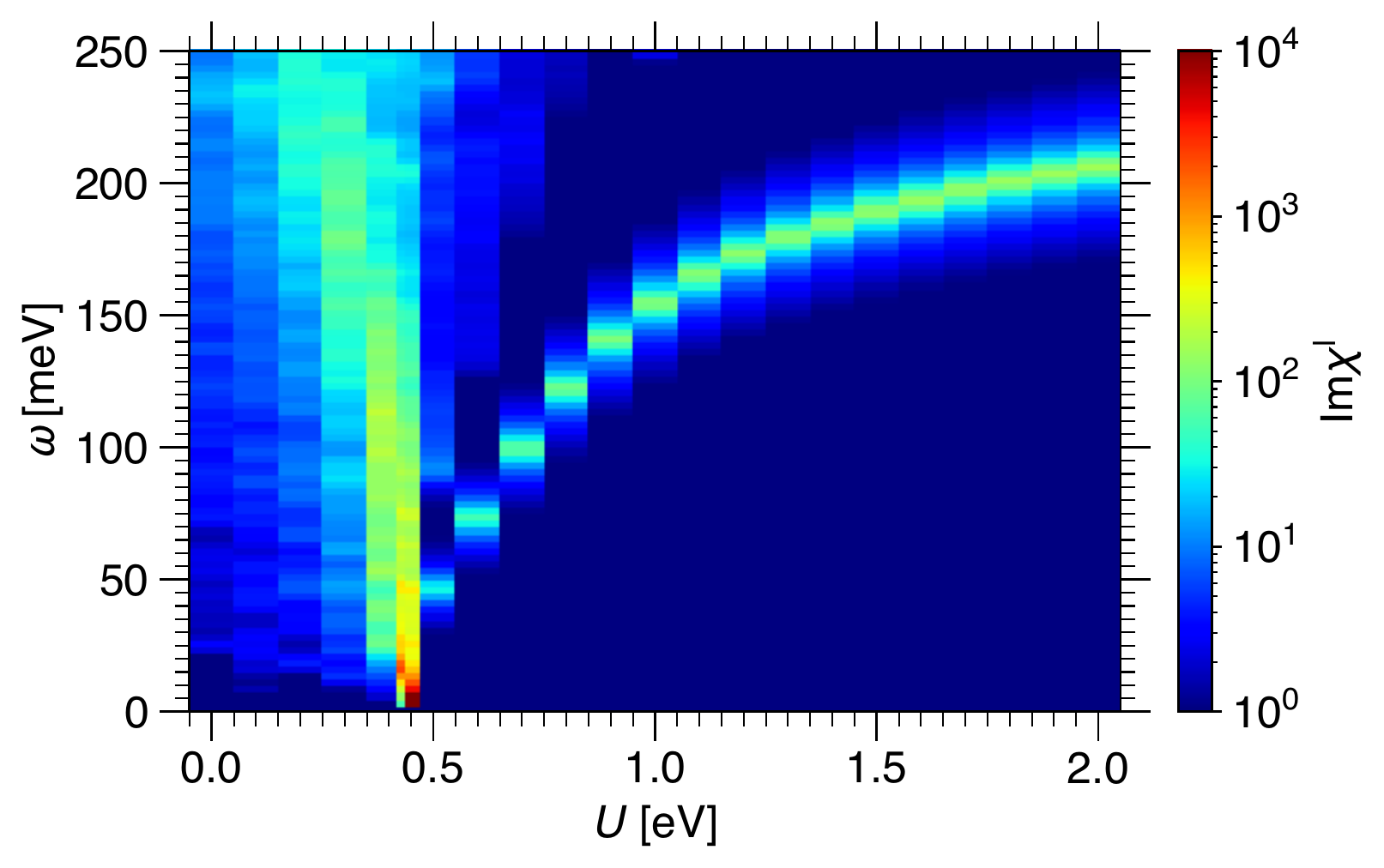}
  \end{center}
  \caption{
    Calculated longitudinal component of the imaginary part of dynamical magnetic susceptibility $\mathrm{Im}~\chi^\mathrm{l}$ at $q = (\pi, \pi)$ as a function of $U$.
    We set $\zeta = 0.15$~eV and the broadening parameter $\eta = 3$~meV.
  }\label{fig:chi_xx_omega_u}
\end{figure}

The characteristic behaviors of the calculated dynamical susceptibility mentioned above are qualitatively consistent with the spectra observed in the previous INS experiments \cite{Kunkemoller2015, Kunkemoller2017, Jain2017NP}, except for the longitudinal mode.
This longitudinal mode is also observed in Raman scattering experiment \cite{Souliou2017}.
In the previous INS experiment, the longitudinal mode appears just above the in-plane transverse mode in its spectra \cite{Jain2017NP}.
However, there is no such dispersion in the present result [see Fig.~\ref{fig:mag_sus}(b)].
To clarify the reason, we investigate the $U$ dependence of the longitudinal component of the susceptibility.
We expect that the longitudinal mode should be gapless at the normal-to-AFM transition point.
For simplicity, we set $t' = 0$ to align the $\bm{S}$ and $\bm{L}$ vectors along the $x$-axis direction.
The result is shown in Fig.~\ref{fig:chi_xx_omega_u}.
We find that the AFM transition occurs at $U \simeq 0.5$~eV, and above the transition point, the newly emerged peak corresponding to the longitudinal mode excitation rapidly grows to a large frequency as $U$ increases.
We, therefore, conclude that the peak position of the longitudinal mode is overestimated due to the mean-field approximation, which becomes comparable to the energy of orbital excitations.
We believe that, although the calculated longitudinal mode is not consistent with the experiment, no problems arise in discussing the RIXS spectra.
This is because the dominant excitation spectra in the RIXS come from the transverse mode, which is well reproduced in our calculations.

\section{RIXS spectrum}
\label{sec:RIXS_spectrum}

In this section, we calculate the RIXS spectra in Ca$_2$RuO$_4$ tuned for the Ru $L_3$ edge.
First, we derive the formula of spectral intensity in the fast-collision approximation, where the dynamical susceptibilities calculated in Sec.~\ref{sec:model_and_method} are included.
Then, the calculated RIXS spectra are analyzed from the viewpoint of collective excitations.

\subsection{Formulation of RIXS spectral intensity}

We briefly introduce the direct RIXS process in ruthenate.
After irradiation of x ray with the Ru $L$-edge frequency, the following three-step process occurs \cite{Ament2011RMP, Ishii2013JPSJ}.
First, the incident photon excites the electron in the core-level $2p$ orbital in Ru atoms to the conduction band.
Next, the electron in the conduction band interacts with the electrons in the valence band.
Finally, the electron in the valence band falls into the hole of the core-level $2p$ orbital, simultaneously emitting photons.
Since the incident photon and emitted photon have a different energy, this scattering process is inelastic.

We now introduce the dipole transition operator $D_{\bm k, \bm\varepsilon}$~($D_{\bm k, \bm\varepsilon}^\dag$), which describes the x-ray absorption~(emission) as
\begin{align}
  D_{\bm k, \bm\varepsilon} = \sum_{\bm k', j, j_z, \mu, \sigma}c^{j, j_z}_{\mu, \sigma}\qty(\bm \varepsilon)c_{\bm k' + \bm k,\mu, \sigma}^\dag p_{\bm k', j, j_z},
\end{align}
where $\bm{k}$ and $\bm{\varepsilon}$ are the wave vector and polarization vector of the x ray, respectively, and $p_{\bm k, j, j_z}$ is the annihilation operator of an electron in the core-level Ru $2p$-orbital with total angular momentum $j$, whose $z$ component is $j_z$.
The matrix element of the dipole operator is given by
\begin{align}
  c_{\mu, \sigma}^{j, j_z}(\bm{\varepsilon})
    = \mel{4d, \mu, \sigma}
    {\bm{\varepsilon} \vdot \bm{r}}
    {2p, j, j_z},
  \label{eq:dipole_operator}
\end{align}
where $\ket{4d, \mu, \sigma}$ and $\ket{2p, j, j_z}$ represent the states with $4d$ and $2p$ orbitals in the Ru atom.
The matrix elements are listed in Appendix~\ref{sec:dipole_operator}.
From the resonance terms of the second-order response of the external field,
the scattering intensity of resonant x ray is given by
\begin{multline}
  I_\mathrm{RIXS}
    (\bm{q} = \bm{k}_\mathrm{in} - \bm{k}_\mathrm{out},
    \omega = \omega_\mathrm{in} - \omega_\mathrm{out}, \bm{\varepsilon}_\mathrm{in}, \bm{\varepsilon}_\mathrm{out})
\\
  \propto \sum_f
    \abs{
      \mel{f}
        {D_{\bm{k}_\mathrm{out}, \bm\varepsilon_\mathrm{out}}^\dag
          \frac{\ketbra{n}}{\omega_\mathrm{in} + E_i - E_n + i \Gamma} D_{\bm{k}_\mathrm{in}, \bm\varepsilon_\mathrm{in}}}
        {i}
    }^2
\\
  \times \delta(\omega - E_f + E_i),
\label{eq:irixs}
\end{multline}
where $E_i$ ($E_f$) is the energy of the initial (final) state, $E_n$ is the energy of the intermediate state, $\bm{\varepsilon}_{\mathrm{in}}$ ($\bm{\varepsilon}_{\mathrm{out}}$) is  polarization vector of incoming (outgoing) x ray, and $1/\Gamma$ represents the lifetime of the intermediate state.
Hereafter, we consider the Ru $L_3$-edge x-ray absorption, i.e., $j = 3/2$.
Furthermore, to calculate the RIXS intensity, we apply the fast-collision approximation; i.e., the lifetime of the intermediate state is assumed to be sufficiently short compared to the scale of the electron motion, so that we neglect the dynamics in the intermediate state.
With this approximation, Eq. (\ref{eq:irixs}) is simplified as
\begin{align}
  I_\mathrm{RIXS}(\bm{q}, \omega, \bm{\varepsilon}_\mathrm{in}, \bm{\varepsilon}_\mathrm{out})
  &\propto \Im \sum_{u, v} \chi_{uv} \qty(\bm{q}, \omega)\notag\\
  &\times \sum_{j_z}c_{\kappa,\sigma_1}^{j, j_z} \qty(\bm{\varepsilon}_{\mathrm{out}})c_{\lambda,\sigma_2}^{j, j_z}\qty(\bm{\varepsilon}_{\mathrm{in}})^*\notag\\
  &\times \sum_{j_z'} c_{\mu,\sigma_3}^{j, j_z'} \qty(\bm{\varepsilon}_{\mathrm{in}}) c_{\nu,\sigma_4}^{j, j_z'} \qty(\bm{\varepsilon}_{\mathrm{out}})^*.
\label{eq:RIXS_intensity}
\end{align}
We assume that the polarization of the outgoing x rays is not taken into account, i.e., the intensity of RIXS spectral is calculated as a sum of the spectra with $\sigma$- and $\pi$-polarized $\bm{\varepsilon}_{\mathrm{out}}$.

\begin{figure}
  \begin{center}
    \includegraphics[width=1.0\columnwidth]{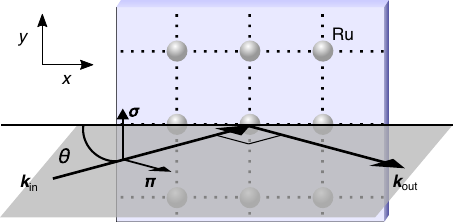}
  \end{center}
  \caption{
      The scattering geometry used in the calculations of RIXS spectra.
      The gray spheres represent Ru atoms.
      We assume that the scattering plane is perpendicular to the square lattice.
      The direction of $\sigma$ ($\pi$) polarization is perpendicular (parallel) to the scattering plane.
  }\label{RIXSgeometory}
\end{figure}

Figure~\ref{RIXSgeometory} illustrates the scattering geometry assumed in the calculation of the RIXS spectra.
$\bm{k}_\mathrm{in}$ ($\bm{k}_\mathrm{out}$) is the wave vector of the incident (scattered) x ray, and we assume that the angle between $\bm{k}_\mathrm{in}$ and $\bm k_\mathrm{out}$ is equal to $\pi / 2$, and the scattering plane is perpendicular to the square lattice.
We denote the angle between the $xy$-plane and $\bm{k}_{\mathrm{in}}$ as $\theta$.
Since we consider the two-dimensional system, the momentum transfer $\bm{q}$ is equal to $\bm{k}_\mathrm{in} - \bm{k}_\mathrm{out}$ projected onto the $xy$-plane.
The energy of the dipole-active $L_3$ edge of Ru atom is about 2838.5~eV \cite{Gretarsson2019}, which corresponds to $\abs{\bm k_\mathrm{in}} \simeq \abs{\bm k_\mathrm{out}} \simeq 1.76\pi$.
Therefore, by varying the angle of the incident x-ray, the momentum transfer can sweep the entire Brillouin zone.
We investigate the RIXS spectra in both low-energy and high-energy regions.

\subsection{Simulated RIXS spectra}

\begin{figure*}
  \begin{tabular}{cccc}
  \begin{minipage}[t]{0.25\hsize}
    \centering
    \includegraphics[height=1.14\columnwidth]{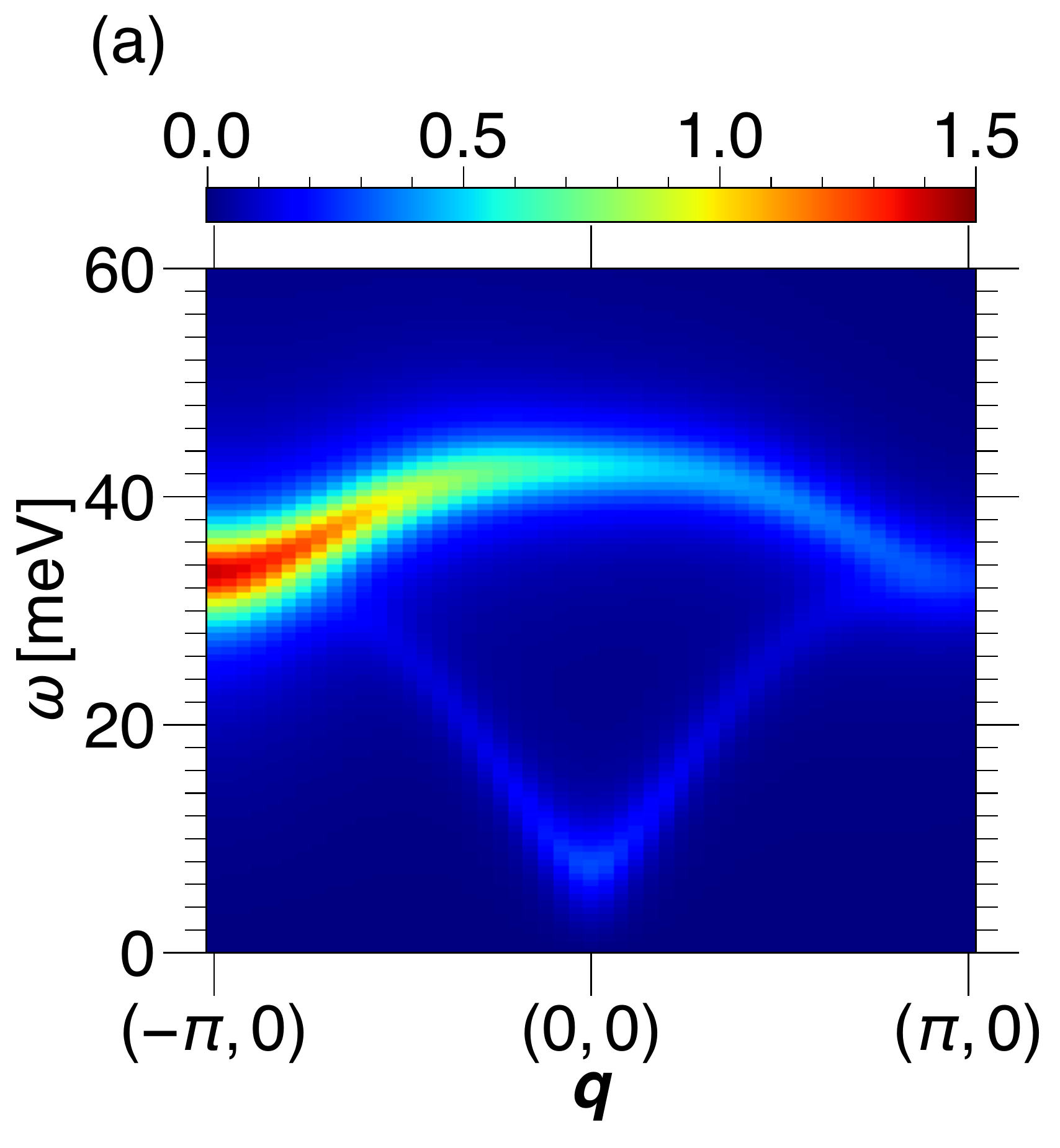}
  \end{minipage}&
  \begin{minipage}[t]{0.25\hsize}
    \centering
    \includegraphics[height=1.14\columnwidth]{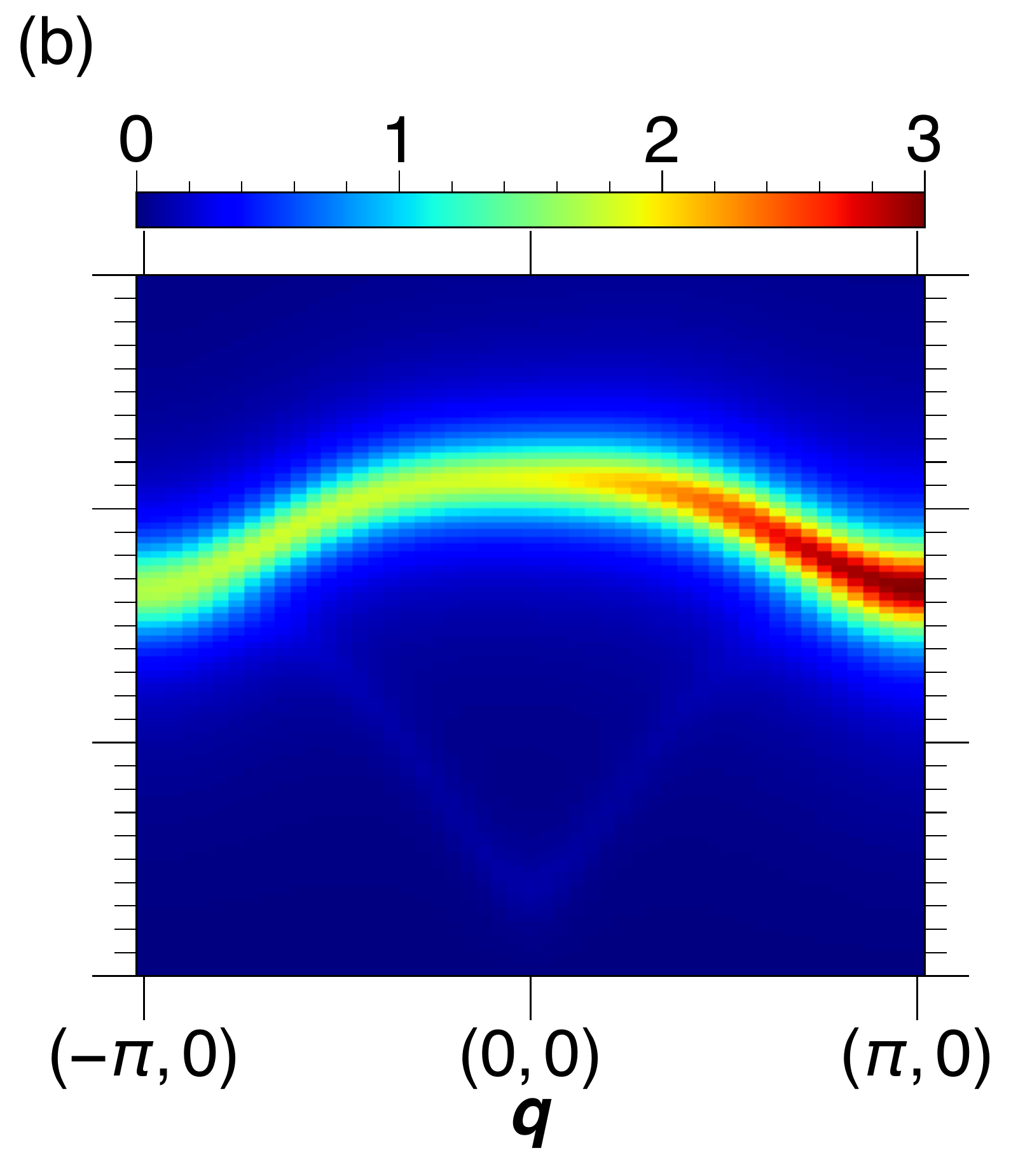}
  \end{minipage}&
  \begin{minipage}[t]{0.25\hsize}
    \includegraphics[height=1.14\columnwidth]{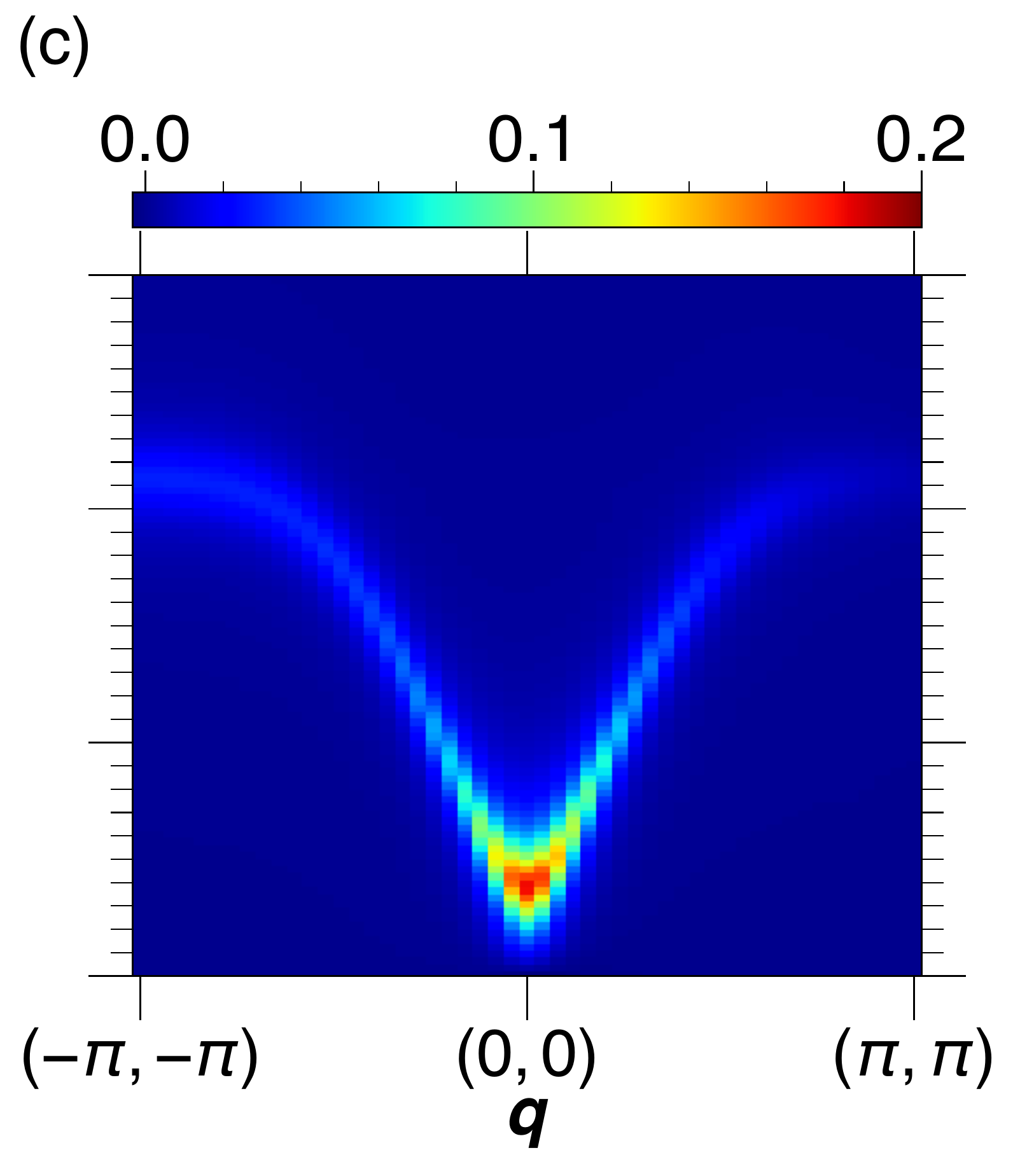}
  \end{minipage}&
  \begin{minipage}[t]{0.25\hsize}
    \includegraphics[height=1.14\columnwidth]{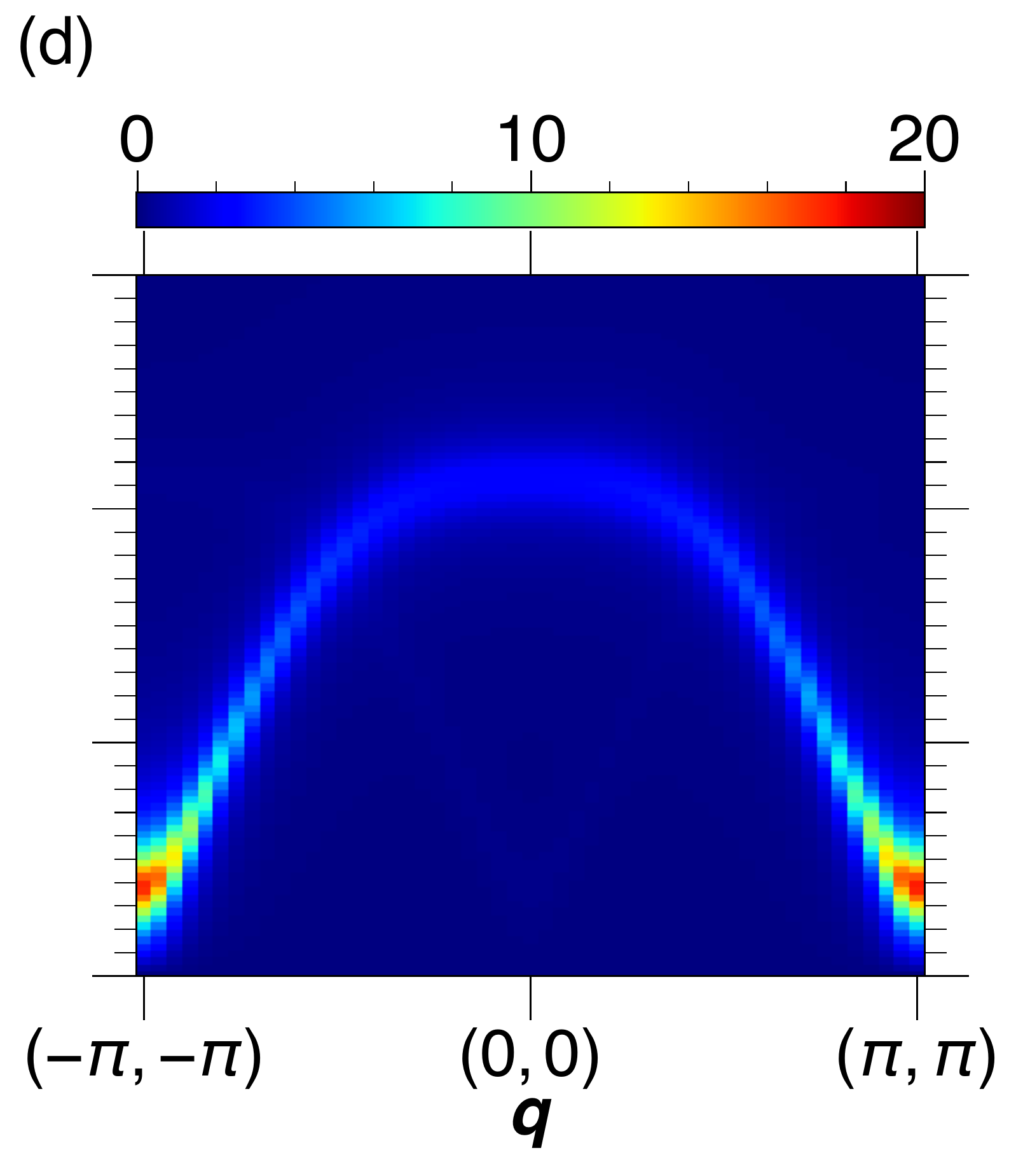}
  \end{minipage}
  \end{tabular}
  \caption{
    Calculated dispersion of the RIXS spectra.
    The momentum transfer is from $\bm q = (-\pi, 0)$ to $(\pi,0)$ in (a) and (b), and is from $\bm q = (-\pi, -\pi)$ to $(\pi, \pi)$ in (c) and (d).
    The incident x rays are $\sigma$ polarized in (a) and (c) and $\pi$ polarized in (b) and (d).
    We set the broadening parameter $\eta = 3$~meV.
  }\label{fig:RIXSdispersion}
\end{figure*}
The RIXS spectra of $\sigma$ and $\pi$ polarization for $\bm q = (-\pi,0)$ to $(\pi, 0)$ with $\bm{k}_\mathrm{in}$ and $\bm{k}_\mathrm{out}$ in the $(k_y=0)$ plane are plotted in  Figs.~\ref{fig:RIXSdispersion}(a) and \ref{fig:RIXSdispersion}(b).
In this case, the angle of incident light is taken from $\theta = 0.382\pi$ to $0.118\pi$.
The results show that when the incident light is $\sigma$ polarized, the strong intensity originating from the in-plane transverse mode appears at ($-\pi, 0$).
On the other hand, when the incident light is $\pi$ polarized, the strong intensity appears at ($\pi, 0$).
We also investigate the RIXS spectra of $\sigma$ and $\pi$ polarization for $\bm q = (-\pi, -\pi)$ to $(\pi, \pi)$ with $\bm{k}_\mathrm{in}$ and $\bm{k}_\mathrm{out}$ in the $(k_x=k_y)$ plane, which are plotted in  Figs.~\ref{fig:RIXSdispersion}(c) and \ref{fig:RIXSdispersion}(d).
In this case, the angle of incident light is taken from $\theta = 0.442\pi$ to $0.058\pi$.
We find that only the out-of-plane transverse mode is observed in the $\sigma$ polarization, while the in-plane transverse mode is observed in the $\pi$ polarization.

\begin{figure}
  \centering
  \includegraphics[height=0.9\columnwidth]{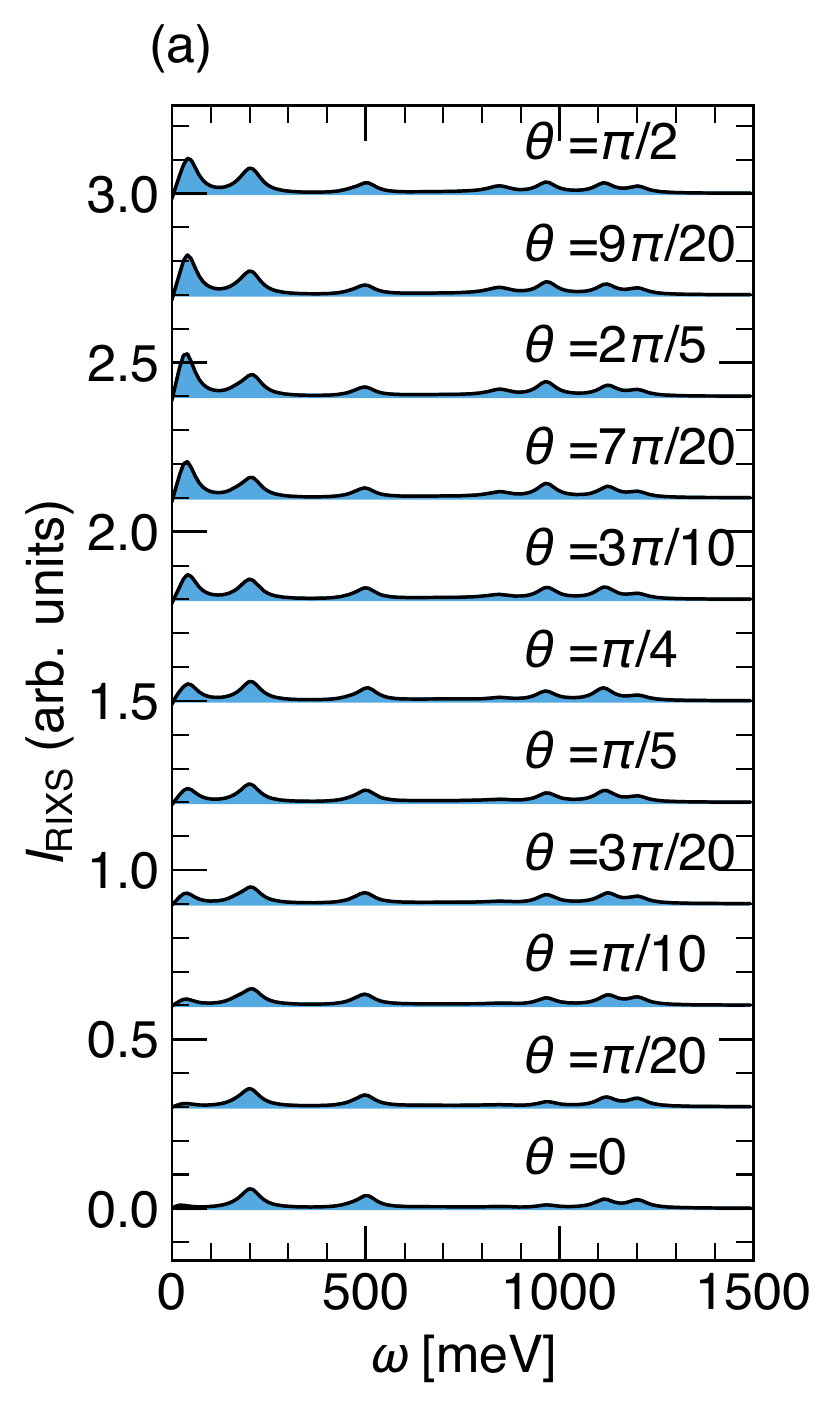}
  \includegraphics[height=0.9\columnwidth]{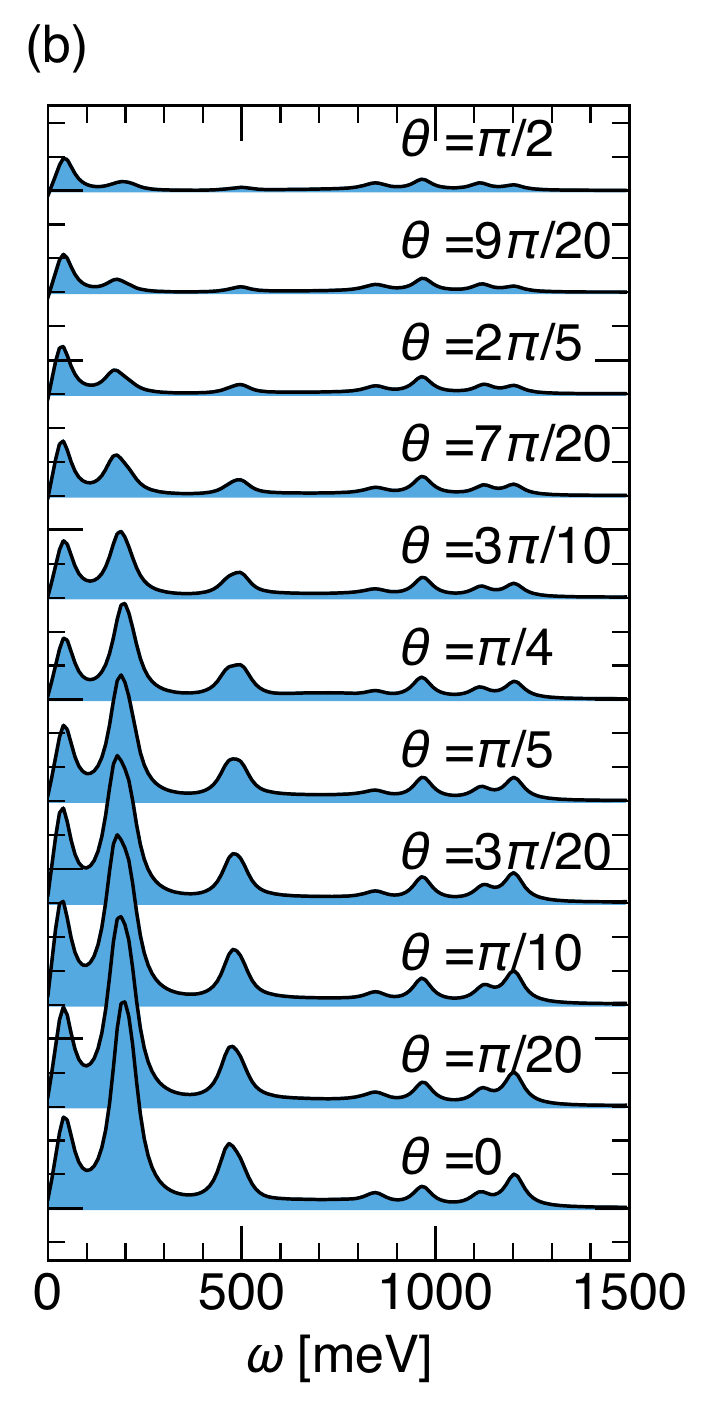}
  \caption{
    Calculated RIXS spectra, where the incoming vector has (a) $\sigma$ poralization and (b) $\pi$ poralization.
    Each line is plotted for the angle of incident light $\theta$ between $0$ and $\pi/2$ from bottom to top with the increments of $\pi/20$.
    We set the broadening parameter $\eta = 30$~meV.
  }\label{fig:RIXStheta}
\end{figure}

Finally, we investigate the RIXS spectra in the high-energy region.
Figure \ref{fig:RIXStheta} shows the RIXS spectra obtained by varying the angle $\theta$ with $\bm{k}_\mathrm{in}$ and $\bm{k}_\mathrm{out}$ in the $(k_y=0)$ plane; the results for $\sigma$ ($\pi$)poralization are shown in Fig.~\ref{fig:RIXStheta}(a) [Fig.~\ref{fig:RIXStheta}(b)].
We find three peaks around 50~meV, 200~meV, and 500~meV.
The 50meV peak corresponds to the collective excitation of the transverse mode discussed in Sec.~\ref{subsec:dynamical_magnetic_susceptibility}.
The 200 (500)~meV peak originates from the excitation from the $d_{xy}$ orbital to $d_{yz/xz}$ orbital with spin conservation (spin flipping) (see Appendix~\ref{sec:spin_orbital_resolved_susceptibility}).
We find that the intensity of the RIXS spectra in the $\pi$ polarization is much larger than that in the $\sigma$ polarization.
In particular, the 200 and 500~meV peaks for $\pi$ polarization show significant intensity at $\theta = 0$, which gradually decrease by increasing $\theta$ from zero to $\pi/2$; the result is consistent with the $\theta$ dependence of the asymmetric peak at 320 meV observed in the experiment \cite{Gretarsson2019}.

Therefore, using RIXS, we can selectively observe the collective mode excitations of Ca$_2$RuO$_4$ by changing the polarization of the incident light.
In other words, the measurement by RIXS has the potential to distinguish a particular kind of collective mode from the excitation spectra.

\section{conclusion}
\label{sec:conclusion}

We have analyzed the collective excitations in \ce{Ca2RuO4} by the itinerant electron approach.
We have introduced the three-orbital Hubbard model with SOC, which is an effective model of Ca$_2$RuO$_4$.
We have applied the mean-field approximation to the model and have obtained the AFM ground state.
Using this state, we have calculated the dynamical magnetic susceptibility by RPA and have obtained the excitation spectra of in-plane and out-of-plane transverse modes.
We thus found that the excitation spectra are consistent with the spectra observed in the previous INS experiments, which confirms the validity of this model.

To calculate the RIXS spectra, we have applied the fast-collision approximation.
Using the dynamical susceptibility calculated in the RPA, we have obtained the Ru $L_3$-edge RIXS spectra of Ca$_2$RuO$_4$.
We have found that the RIXS spectra are quite asymmetric concerning momentum transfer $\bm k$ and $-\bm{k}$ and that the in-plane and out-of-plane transverse modes can be distinguished by varying the polarization of incident light.

The results obtained in this paper are consistent with the experimental results, except for the longitudinal mode observed in the INS spectra.
This may be due to the fact that the mean-field approximation overestimates the magnitude of AFM order, so that the peak position of the longitudinal-mode excitation appears higher in energy than expected.
To obtain the longitudinal mode with appropriate excitation energy in the itinerant electron approach, it is necessary to go beyond the mean-field approximation and RPA, incorporating the quantum fluctuations driven by electron-electron correlations more accurately.
However, we stress that our results properly reproduce the excitation spectra of the transverse mode and that the predicted RIXS spectra can advance our understanding of this material, which we hope will lead to a better understanding of the character of collective excitations in strong SOC materials in general.
We expect that the selective behavior for the polarization of incident light in the RIXS spectra will be observed experimentally in the future.

\section*{Acknowledgments}
We thank H. Fukazawa, T. Yamaguchi and R. Fujiuchi for enlightening discussions.
This work was supported in part by Grants-in-Aid for Scientific Research from JSPS
(Projects No. JP17K05530, No. JP19K14644, and No. JP20H01849).
S.Y. acknowledges the support by JST, the establishment of University fellowships towards the creation of science technology innovation, Grant No. JPMJFS2107.

\appendix

\section{List of matrix elements of dipole operator}
\label{sec:dipole_operator}
\begin{table}
  \centering
  \caption{$c_{\mu,\sigma}^{j, j_z}(\varepsilon_\alpha)$ of the $t_{2g}$ basis with $j=3/2$ and $1/2$.}
  \label{dipoleele}
  \begin{tabular}{clcccccc}
  \hline
  &&\multicolumn{4}{c}{$j=3/2$}&\multicolumn{2}{c}{$j=1/2$}\\
  \cmidrule(lr){3-6}
  \cmidrule(lr){7-8}
  $\alpha$&$\qty(\mu,\sigma)$\,\textbackslash\,$j_z$
  &$3/2$
  &$1/2$
  &$-1/2$
  &$-3/2$
  &$1/2$
  &$-1/2$\\
  \hline
  $x$&$\qty(xz,\uparrow)$&---&$\sqrt{2/15}$&---&---&$1/\sqrt{15}$&---\\
  &$\qty(xz,\downarrow)$&---&---&$\sqrt{2/15}$&---&---&$-1/\sqrt{15}$\\
  &$\qty(xy,\uparrow)$&$-i/\sqrt{10}$&---&$-i/\sqrt{30}$&---&---&$-i/\sqrt{15}$\\
  &$\qty(xy,\downarrow)$&---&$-i/\sqrt{30}$&---&$-i/\sqrt{10}$&$i/\sqrt{15}$&---\\
  $y$&$\qty(yz,\uparrow)$&---&$\sqrt{2/15}$&---&---&$1/\sqrt{15}$&---\\
  &$\qty(yz,\downarrow)$&---&---&$\sqrt{2/15}$&---&---&$-1/\sqrt{15}$\\
  &$\qty(xy,\uparrow)$&$-1/\sqrt{10}$&---&$1/\sqrt{30}$&---&---&$1/\sqrt{15}$\\
  &$\qty(xy,\downarrow)$&---&$-1/\sqrt{30}$&---&$1/\sqrt{10}$&$1/\sqrt{15}$&---\\
  $z$&$\qty(yz,\uparrow)$&$-i/\sqrt{10}$&---&$-i/\sqrt{30}$&---&---&$-i/\sqrt{15}$\\
  &$\qty(yz,\downarrow)$&---&$-i/\sqrt{30}$&---&$-i\sqrt{10}$&$i/\sqrt{15}$&---\\
  &$\qty(xz,\uparrow)$&$-1/\sqrt{10}$&---&$1/\sqrt{30}$&---&---&$1/\sqrt{15}$\\
  &$\qty(xz,\downarrow)$&---&$-1/\sqrt{30}$&---&$1/\sqrt{10}$&$1/\sqrt{15}$&---\\
  \hline
  \end{tabular}
\end{table}
To obtain the intensity of the RIXS spectra, we should calculate the matrix elements of the dipolar operator given in Eq.~(\ref{eq:dipole_operator}).
If we write the polarization vector as $\bm{\varepsilon} = (\varepsilon_x, \varepsilon_y, \varepsilon_z)$, then the matrix element in $\ket{n,l,m}$ basis, where $n$, $l$, and $m$ are principal, azimuthal, and magnetic quantum numbers, is given by
\begin{widetext}
\begin{multline}
  \langle n', l', m' | \bm{\varepsilon} \cdot \bm{r} | n, l, m \rangle
   = \sqrt{\frac{4\pi}{3}} \int^{\infty}_{0} \dd{r} r^3 R_{n' l'}^* (r) R_{nl} (r) \\
   \times
   \qty[
     \frac{-\varepsilon_x + i \varepsilon_y}{\sqrt{2}}  c^1 (l', m+1; l, m) \delta_{m', m+1}
     + \frac{\varepsilon_x + i \varepsilon_y}{\sqrt{2}} c^1 (l', m-1; l, m) \delta_{m', m-1}
     + \varepsilon_z c^1 (l', m; l, m) \delta_{m',m}
   ],
 \label{eq:dipoleele}
 \end{multline}
 \end{widetext}
 where $R_{nl} (r)$ is the radial wave functions of a hydrogen atom and $c^{l_1} (l\rq, m\rq ; l_2, m_2)$ is the Gaunt coefficient defined as the integral over three spherical harmonics.
To obtain the $L$ edge scattering amplitude for Ru atoms, we have to calculate the dipole matrix
element between $2p$ and $t_{2g}$ of $4d$ orbitals.
By transforming the basis of Eq. (\ref{eq:dipoleele}), we can obtain such matrix elements as
\begin{align}
  c_{\mu,\sigma}^{j, j_z}(\varepsilon_\alpha) = \mel{\mu,\sigma}{\varepsilon_\alpha \cdot r_\alpha}{j,j_z},
\end{align}
whose specific values are listed in Table \ref{dipoleele}.

\section{Spin-orbital resolved susceptibility}
\label{sec:spin_orbital_resolved_susceptibility}
\begin{figure}
  \begin{center}
    \includegraphics[width=1.0\columnwidth]{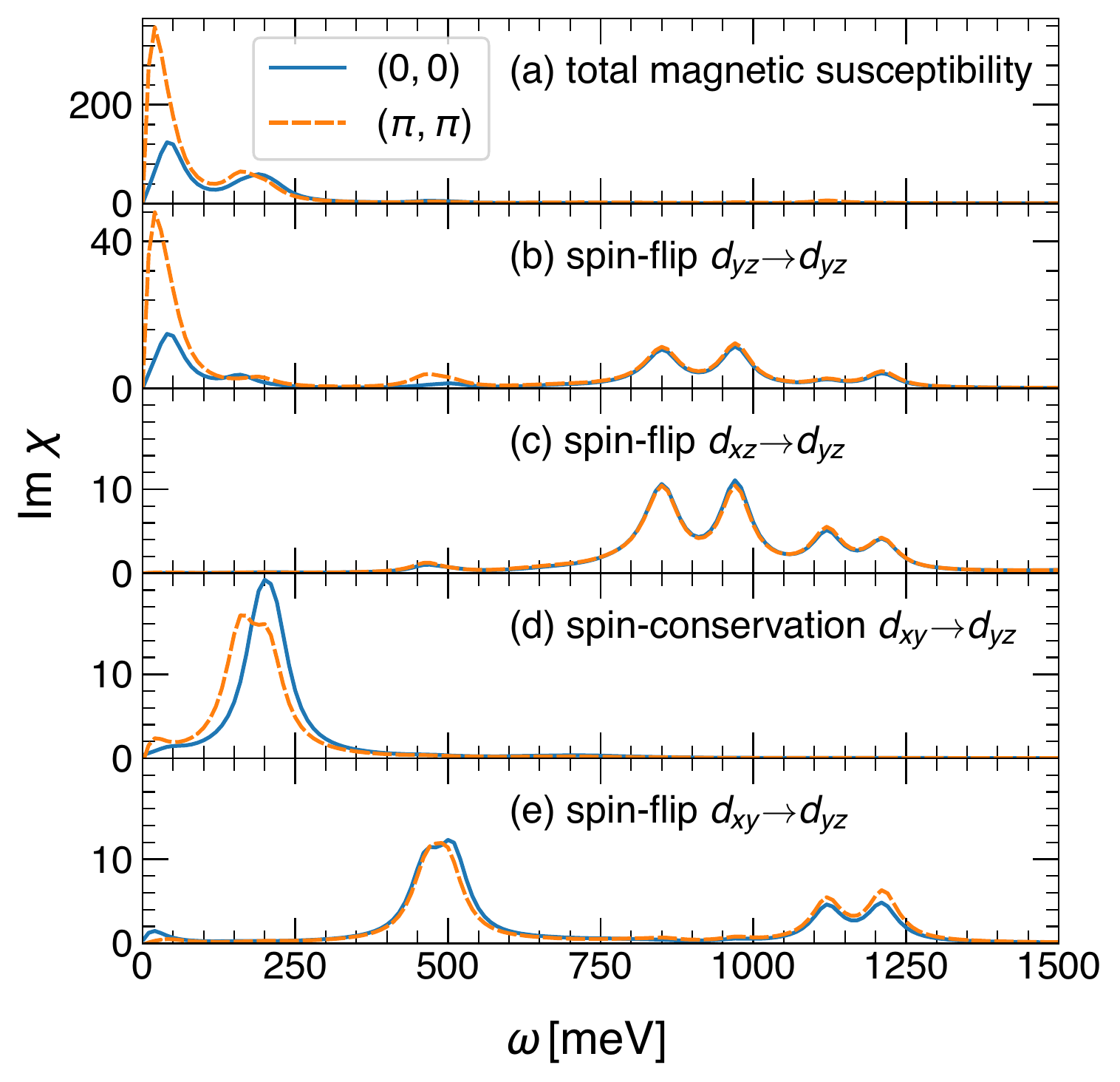}
  \end{center}
  \caption{Calculated susceptibilities at $\bm q = (0, 0)$ and $\bm q = (\pi, \pi)$. (a) Total magnetic susceptibility. (b) Spin-flip susceptibility in $d_{yz}$ orbital. (c) Spin-flip susceptibility between $d_{yz}$ and $d_{xz}$ orbitals. (d) Spin-conservation susceptibility between $d_{yz}$ and $d_{xy}$ orbitals. (e) Spin-flip susceptibility between $d_{yz}$ and $d_{xy}$ orbitals.
  We set the broadening paramete $\eta = 30$~meV.
  }\label{orbital_resolved_suceptibility}
\end{figure}
We investigated the spin-orbital resolved susceptibility to interpret the excitation spectra.
Figure~\ref{orbital_resolved_suceptibility}(a) shows the total magnetic susceptibility, showing the low-energy ($<50$~meV) peaks originating from the transverse mode and high-energy ($\sim 200$~meV) peaks.
Next, we calculated spin-orbital resolved susceptibility, which is given by
\begin{align}
  \chi \qty(\bm{q}, \omega)
    = \frac{i}{N} \int_0^\infty \dd{t}
      e^{i\omega t}\ev{\comm{O^{\dag}_{\bm{q}} (t)}{O_{-\bm{q}}(0)}},
\end{align}
with
\begin{align}
    O_{\bm{q}} = \sum_{\bm k} \bm{c}_{\bm{k}}^\dag (O_3\otimes O_2) \bm{c}_{\bm{k}+\bm{q}},
\end{align}
where $O_{3(2)}$ is a matrix in orbital (spin) space.

Figure~\ref{orbital_resolved_suceptibility}(b) shows the spin-flip susceptibility in the $d_{yz}$ orbital with
\begin{align}
  O_3\otimes O_2 = \mqty(1&0&0\\0&0&0\\0&0&0)\otimes \mqty(1&i\exp(i\frac{\pi}{4})\\i\exp(-i\frac{\pi}{4})&-1).
\end{align}
This susceptibility contributes to transverse mode ($<50$~meV).

Figure~\ref{orbital_resolved_suceptibility}(c) shows the spin-flip susceptibility between the $d_{yz}$ and $d_{xz}$ orbital with
\begin{align}
  O_3\otimes O_2 = \mqty(0&1&0\\0&0&0\\0&0&0)\otimes\mqty(1&i\exp(i\frac{\pi}{4})\\i\exp(-i\frac{\pi}{4})&-1).
\end{align}
This susceptibility does not contribute to the low-energy excitation.

Figure~\ref{orbital_resolved_suceptibility}(d) shows the spin-conservation susceptibility between the $d_{yz}$ and $d_{xy}$ orbitals with
\begin{align}
  O_3\otimes O_2 = \mqty(0&0&1\\0&0&0\\0&0&0)\otimes \mqty(1&0\\0&1).
\end{align}
This susceptibility has a peak at 200~meV.

Figure~\ref{orbital_resolved_suceptibility}(e) shows the spin-flip susceptibility between the $d_{yz}$ and $d_{xy}$ orbitals with
\begin{align}
  O_3\otimes O_2 = \mqty(0&0&1\\0&0&0\\0&0&0)\otimes \mqty(1&i\exp(i\frac{\pi}{4})\\i\exp(-i\frac{\pi}{4})&-1).
\end{align}
This susceptibility has a peak at 500~meV, which does not contribute to the INS spectrum, but contributes to the RIXS spectrum (see Fig.~\ref{fig:RIXStheta}).
\bibliographystyle{apsrev4-1}
%
\end{document}